\documentclass[twocolumn,secnumarabic,showpacs,amssymb, amsmath,tightenlines,aps,prb]{revtex4}
\usepackage{longtable}
\usepackage{bm}
\usepackage{epsfig}
\usepackage{graphicx}

\begin{document}

\title[High-field moment polarization in the ferromagnetic superconductor UCoGe; similarities with the URhGe-case]{High-field moment polarization in the ferromagnetic superconductor UCoGe; similarities with the URhGe-case}
\author{W. Knafo$^{1}$, T.D. Matsuda$^{2}$, D. Aoki$^{3,4}$, F. Hardy$^{5}$, G.W. Scheerer$^{1}$, G. Ballon$^{1}$, M. Nardone$^{1}$, A. Zitouni$^{1}$, C. Meingast$^{5}$, and J. Flouquet$^{3}$}

\address{$^{1}$ Laboratoire National des Champs Magn\'{e}tiques Intenses, UPR 3228, CNRS-UJF-UPS-INSA, 143 Avenue de Rangueil,
31400 Toulouse, France.\\
$^{2}$ Advanced Science Research Center, Japan Atomic Energy Agency, Tokai, Ibaraki 319-1195, Japan.\\
$^{3}$ Institut Nanosciences et Cryog\'{e}nie, SPSMS, CEA-Grenoble, 17 rue des Martyrs, 38054 Grenoble, France.\\
$^{4}$ Institute for Materials Research, Tohoku University, Oarai, Ibaraki, 311-1313, Japan.\\
$^{5}$ Institut f\"{u}r Festk\"{o}rperphysik, Karlsruher Institut f\"{u}r Technologie, 76021 Karlsruhe, Germany.}

\date{\today}

\begin{abstract}

We report magnetization and magnetoresistivity measurements on the isostructural ferromagnetic superconductors UCoGe and URhGe in magnetic fields up to 60~T and temperatures from 1.5 to 80~K. At low-temperature, a moment polarization in UCoGe in a field $\mu_0\mathbf{H}\parallel\mathbf{b}$ of around 50~T leads to well-defined anomalies in both magnetization and magnetoresistivity. These anomalies vanish in temperatures higher than 30-40~K, where maxima in the magnetic susceptibility and the field-induced variation of the magnetoresistivity are found. A comparison is made between UCoGe and URhGe, where a moment reorientation in a magnetic field $\mu_0\mathbf{H}\parallel\mathbf{b}$ of 12~T leads to field-induced reentrant superconductivity.

\end{abstract}

\pacs{71.27.+a,74.70.Tx,75.30.Kz}



\maketitle

Field-induced transitions to a paramagnetic polarized regime have been observed in many heavy-fermion systems close to a paramagnetic-to-antiferromagnetic quantum instability. In heavy-fermion paramagnets, a simple correspondence 1~K~$\leftrightarrow$~1~T often holds between the temperature $T_\chi^{max}$ at the maximum of the magnetic susceptibility and the transition field $H_m$, which indicates that they are both controlled by a unique phenomenon \cite{aoki12b}. In the textbook heavy-fermion paramagnet CeRu$_2$Si$_2$, they are related to antiferromagnetic fluctuations, which are quenched at the metamagnetic transition $H_m=8$~T [\onlinecite{sato01,raymond99,flouquet04}] and whose zero-field energy scale, the quasielastic linewitdh at the antiferromagnetic wavevector \cite{knafo09}, equals $T_\chi^{max}=10$~K. Field-induced superconductivity was recently discovered in the ferromagnetic superconductor URhGe [\onlinecite{aoki01,levy05}] in a field $\mathbf{H}$ applied along the hard magnetic axis $\mathbf{b}$, in the window $8\leq \mu_0H\leq12$~T far above the superconducting critical field $\mu_0H_{c,2}=2$~T. In this system, reentrant superconductivity is intimately connected to a moment reorientation occurring at 12~T [\onlinecite{levy05,miyake09}]. In the other ferromagnetic superconductors UCoGe at ambient pressure \cite{huy07,aoki09} and UGe$_2$ under pressure \cite{saxena00,sheikin01}, an anomalous $S$-shape of $H_{c,2}$ could also result from field-induced reentrant superconductivity. The challenge is now to discover a second case where superconductivity develops well above $H_{c,2}$, i.e., with two well-separated superconducting phases, showing that reentrant superconductivity in URhGe is not an isolated exotic case. With this aim, a prerequisite is to find another case of field-induced moment polarization in a superconducting ferromagnet, where ferromagnetic fluctuations could drive to field-induced superconductivity. However, in such complex strongly correlated electron systems, a large field-induced magnetic polarization can also modify the local fluctuations, which are partly responsible for the mass enhancement, and may lead to strong modifications of the Fermi surface. Lifshitz-like transitions were recently proposed to occur in URhGe [\onlinecite{yelland11}] and UCoGe [\onlinecite{malone12}].

Here, we present a study per magnetization and magnetoresistivity of the ferromagnetic superconductor UCoGe in pulsed magnetic fields up to 60~T applied along the three main directions $\mathbf{a}$, $\mathbf{b}$, and $\mathbf{c}$, in the orthorhombic structure (TiNiSi-type). Temperatures from 1.5 to 80~K, i.e., above the zero-field superconducting temperature $T_{sc}=0.6$~K, were investigated. We find that a field-induced moment polarization occurs in a magnetic field of  $\simeq50$~T applied along the hard axis $\mathbf{b}$. This transition vanishes at temperatures higher than 30-40~K, which also corresponds to maxima in the magnetic susceptibility and in the field-dependence of the magnetoresistivity. We emphasize the striking similarity between the moment polarization found here for UCoGe and the moment reorientation occurring at 12~T - for $\mathbf{H}\parallel\mathbf{b}$ too - in URhGe, whose high-field magnetization and resistivity are also studied.

High-quality UCoGe and URhGe single crystals were grown by the Czochralski method in a tetra-arc furnace. Magnetic susceptibility was measured using a commercial SQUID magnetometer (Magnetic Properties Measurement System from Quantum Design). Magnetization of URhGe up to 14~T was measured using a commercial VSM insert in a PPMS (Physical Properties Measurement System from Quantum Design). Experiments in pulsed magnetic fields were performed using standard 150-ms (30-ms rise, 120-ms fall) 60-T magnets of the Laboratoire National des Champs Magn\'{e}tiques Intenses (LNCMI) in Toulouse. High-field magnetization was measured using the compensated-coil technique. High-field magnetoresistivity was measured within the four-point technique at a frequency of $\simeq40$ kHz, using a digital lock-in (developed at the LNCMI by E. Haanappel), and a transverse configuration with the electrical current $\mathbf{I}\parallel\mathbf{c}$ and the magnetic field $\mathbf{H}\parallel\mathbf{b}$. The UCoGe and URhGe samples studied by resistivity had residual resistivity ratios $\rho_{z,z}(300\rm{ K})/\rho_{z,z}^0\simeq45$ and 11, respectively.

\begin{figure}[t]
    \centering
    \epsfig{file=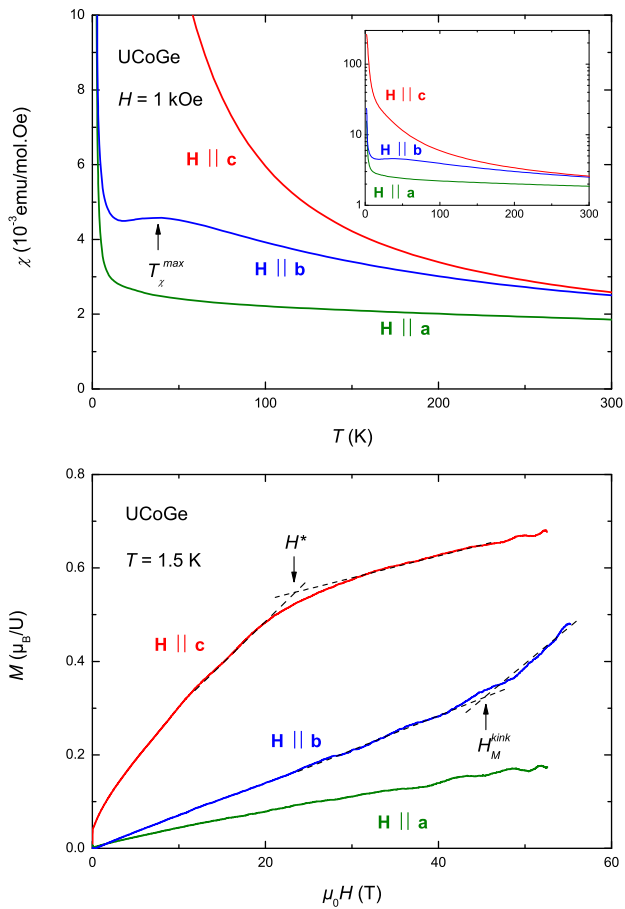,width=83mm}
    \caption{(color online) (a) Magnetic susceptibility versus temperature at $H=1$~kOe (in an extended scale in the Inset) and (b) magnetization versus magnetic field at $T=1.5$~K of UCoGe for $\mathbf{H}\parallel\mathbf{a}$, $\mathbf{b}$, and $\mathbf{c}$.}
    \label{UCoGe_M_H_abc}
\end{figure}

\begin{figure}[t]
    \centering
    \epsfig{file=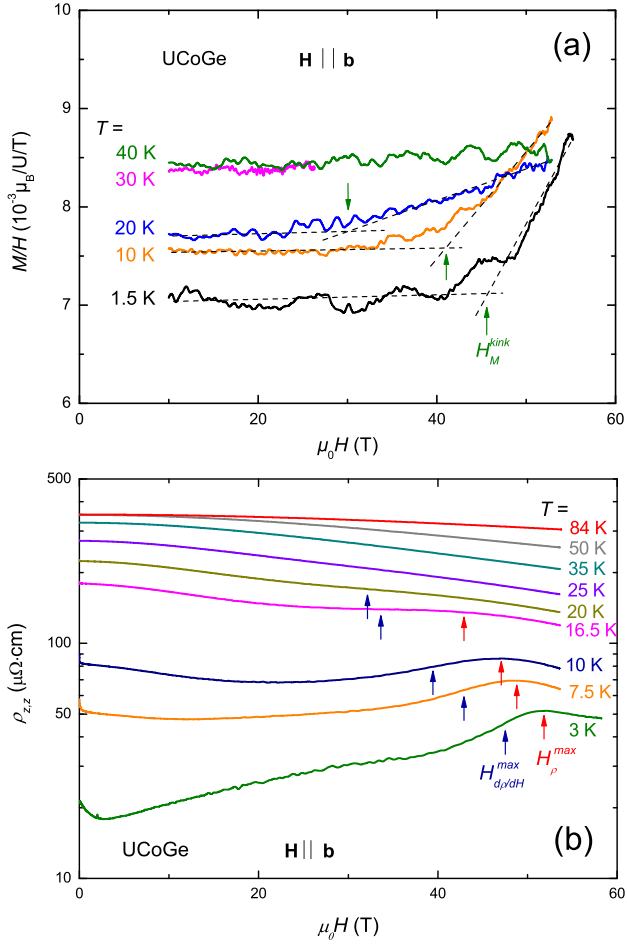,width=83mm}
    \caption{(color online) (a) Magnetization divided by the magnetic field and (b) magnetoresistivity versus field, in a semi-log scale, of UCoGe for $\mathbf{H}\parallel\mathbf{b}$ and at temperatures from 1.5 to 84~K.}
    \label{UCoGe_M_H_T}
\end{figure}

Figure \ref{UCoGe_M_H_abc} (a) presents the magnetic susceptibility versus temperature of UCoGe in a magnetic field $\mathbf{H}$ of 1~kOe applied along $\mathbf{a}$, $\mathbf{b}$, and $\mathbf{c}$, while Fig. \ref{UCoGe_M_H_abc} (b) shows its magnetization versus magnetic field at $T=1.5$~K in $\mu_0\mathbf{H}\parallel\mathbf{a}$, $\mathbf{b}$, and $\mathbf{c}$ up to 53~T. In agreement with previous magnetization experiments up to 5~T [\onlinecite{huy08}], the hierarchies $\chi_c>\chi_b>\chi_a$ and $M_c>M_b>M_a$ indicate that UCoGe has a strong Ising anisotropy, $c$ being the easy magnetic axis, $b$ an intermediate hard magnetic axis, and $a$ the hardest magnetic axis. For $\mathbf{H}\parallel\mathbf{c}$, the slope of $M(H)$ decreases with increasing field, a kink being observed at $\mu_0H^*=23.5\pm2$~T, above which the system is partially polarized; at $\mu_0H=40$~T, $M=0.6$~$\mu_B$/U is still 6 times smaller than the U$^{3+}$ and U$^{4+}$ free ions values 3.62 and 3.58~$\mu_B$/U, respectively, and $M(H)$ continues to increase, indicating the presence of still-unquenched magnetic fluctuations and/or a crystal-field splitting. No anomaly is observed up to 53 T in the magnetization measured for $\mathbf{H}\parallel\mathbf{a}$. For $\mathbf{H}\parallel\mathbf{b}$, a significant enhancement of the slope of $M(H)$ is observed at $\mu_0H_M^{kink}=46\pm2$~T, indicating another field-induced moment polarization. In Fig. \ref{UCoGe_M_H_T} (a), a plot of $M/H$ versus $H$ at different temperatures shows that the anomaly at $H_M^{kink}$ can be followed up to 20~K, its trace being lost above 30~K. This temperature scale coincides with the temperature $T_\chi^{max}\simeq37.5\pm1$~K at which, for $\mathbf{H}\parallel\mathbf{b}$, the magnetic susceptibility of UCoGe is maximal. The magnetic field-dependence, for $\mathbf{H}\parallel\mathbf{b}$ and in a semi-log scale, of the resistivity $\rho_{z,z}$ of UCoGe is presented in Fig. \ref{UCoGe_M_H_T} (b) at different temperatures from 3 to 84~K.  At $T=3$~K, $\rho_{z,z}$ is characterized by a maximum at $\mu_0H_{\rho}^{max}=52\pm0.5$~T. When the temperature increases, a contribution with a negative slope develops in $\rho_{z,z}(H)$, in addition to the anomaly associated to $H_{\rho}^{max}$, which shifts to the lower fields. Above 15~K, $\rho_{z,z}(H)$ decreases monotonously and it is impossible to define $H_{\rho}^{max}$, although a related shoulder is still clearly visible up to 20~K. We can follow the evolution of the high-field anomaly up to 20~K by considering the field $H_{\partial\rho_{z,z}/\partial H}^{max}$ defined at the maximum of $\partial\rho_{z,z}/\partial H$, instead of $H_{\rho}^{max}$. In agreement with the magnetization, the resistivity of UCoGe under $\mathbf{H}\parallel\mathbf{b}$ confirms thus the presence of an anomaly at around 50~T, which vanishes in temperatures higher than 30~K. The temperature scale of 30~K also corresponds to the strongest variation, of negative slope, observed in $\rho_{z,z}(H)$ up to 60~T.

\begin{figure}[t]
    \centering
    \epsfig{file=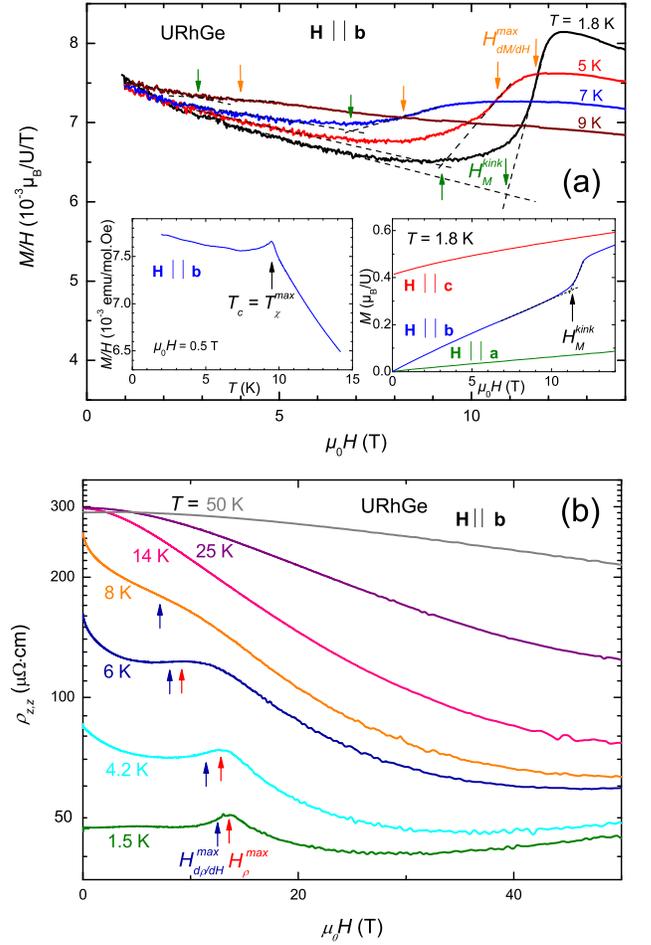,width=83mm}
    \caption{(color online) (a) Magnetization divided by the magnetic field and (b) magnetoresistivity versus field, in a semi-log scale, of URhGe for  $\mathbf{H}\parallel\mathbf{b}$ and at temperatures from 1.5 to 50~K. The insets of (a) show, on the left, the magnetic susceptibility of URhGe versus temperature for $\mu_0\mathbf{H}\parallel\mathbf{b}$ of 0.5~T and, on the right, the magnetization of URhGe versus magnetic field for $\mathbf{H}\parallel\mathbf{a}$, $\mathbf{b}$, and $\mathbf{c}$ and at $T=1.8$~K.}
    \label{URhGe_MsH_Mabc_rho}
\end{figure}

For comparison with the UCoGe case, we present in Fig. \ref{URhGe_MsH_Mabc_rho} the high-field magnetization and magnetoresistivity of URhGe. URhGe is isostructural and has a similar magnetic anisotropy than UCoGe, with $M_c>M_b>M_a$ [cf. right Inset of Fig. \ref{URhGe_MsH_Mabc_rho} (a)]. As shown in Fig. \ref{URhGe_MsH_Mabc_rho} (a) (see also Ref. [\onlinecite{levy05}]) we observe in $M(H)$, for $\mathbf{H}\parallel\mathbf{b}$ and at $T=1.8$~K, a step-like anomaly which can be defined either at the kink ($\mu_0H_M^{kink}=10.9\pm0.5$~T) or at the maximum of slope of $M(H)$ ($\mu_0H_{\partial M/\partial H}^{max}=11.7\pm0.1$~T). This anomaly moves to the lower fields when $T$ is increased and it vanishes above $T_C=9.5$~K. A sharp maximum is observed at $T_\chi^{max}=T_C$ in the magnetic susceptibility of URhGe for $\mathbf{H}\parallel\mathbf{b}$ [see left Inset of Fig. \ref{URhGe_MsH_Mabc_rho} (a)]. Figure \ref{URhGe_MsH_Mabc_rho} (b) shows the magnetic field-dependence, in a semi-log scale, of the resistivity $\rho_{z,z}$ of URhGe for $\mathbf{H}\parallel\mathbf{b}$, at temperatures from 1.5 to 50~K. A maximum in the resistivity is observed at $\mu_0H_{\rho}^{max}=13.6\pm0.5$~T [\onlinecite{noteURhGe}], which decreases when $T$ is increased. For comparison with UCoGe [cf. Fig. \ref{UCoGe_M_H_abc} (b)], the field $\mu_0H_{\partial\rho_{z,z}/\partial H}^{max}=12.5\pm0.5$~T at the maximum of slope of $\partial\rho_{z,z}/\partial H$ is also extracted. The trace of the high-field anomaly is lost for $T>T_C$. An additional contribution (negative slope) to $\rho_{z,z}(H)$ also develops at low-field when $T$ is increased, leading to a maximal variation at 10~K of $\rho_{z,z}(H)$, in a window up to 60~T.

\begin{figure}[t]
    \centering
    \epsfig{file=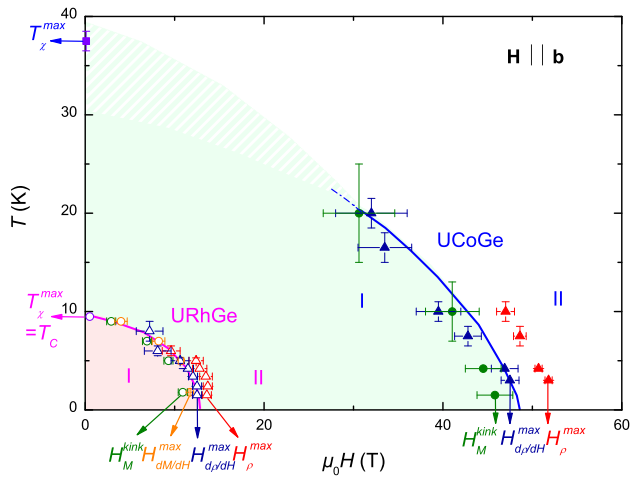,width=83mm}
    \caption{(color online) Magnetic field - temperature phase diagram of UCoGe and URhGe extracted here from magnetization and magnetoresistivity measurements with $\mathbf{H}\parallel\mathbf{b}$. For each compound, I and II indicate the regimes below and above the moment polarization field.}
    \label{ph_diag}
\end{figure}

Figure \ref{ph_diag} shows the magnetic field - temperature phase diagrams, with $\mathbf{H}\parallel\mathbf{b}$, extracted from our magnetization and magnetoresistivity measurements on UCoGe and URhGe. The characteristic field at the moment polarization and the temperature $T_\chi^{max}$ at the maximum of the susceptibility are plotted for the two compounds. Since the anomalies are rather broad (in particular for UCoGe), their characteristic field is difficult to define and we extract it using different criteria: $H_M^{kink}$ and $H_{\partial M/\partial H}^{max}$ from the magnetization and $H_{\rho}^{max}$ and $H_{\partial\rho_{z,z}/\partial H}^{max}$ from the magnetoresistivity. In UCoGe, the anomaly at 50~T observed at low-temperature vanishes for $T>30$~K ($\simeq T_\chi^{max}$). As well, the anomaly at 12~T observed in URhGe at low-temperature vanishes for $T>9.5$~K ($=T_\chi^{max}$). From the Maxwell relation $(\partial S/\partial H)_T=(\partial M/\partial T)_H$ the magnetic entropy and thus the magnetic fluctuations increase with $H$ when $T<T_{\chi}^{max}$, indicating the proximity of a field-induced transition \cite{BealMonod1982}. Table \ref{table} summarizes the characteristic temperature scales, the spontaneous moment $M_s$, and the moment polarization field, noted here $H_{m}$, of URhGe and UCoGe for $\mathbf{H}\parallel\mathbf{b}$. While the spontaneous moment $M_s$ of UCoGe (0.05~$\mu_B$/U) is much smaller than that of URhGe (0.4~$\mu_B$/U), the magnetization of both systems reaches similar values ($\simeq0.6$~$\mu_B$/U) in high fields $\mathbf{H}\parallel\mathbf{b},\mathbf{c}$. Another striking similarity is that, for $\mathbf{H}\parallel\mathbf{b}$, the moment polarization occurs in both UCoGe and URhGe when $M$ reaches $\simeq0.3\mu_B$/U. In URhGe, $M_s=0.4$~$\mu_B$/U is large and the field $H_m$ coincides with $M\simeq M_s$, indicating that the polarization of the ferromagnetic moments at $H_m$ has a feedback on superconductivity. In UCoGe, $M_s=0.05$~$\mu_B$/U is small and $H_m$ corresponds to $M\gg M_s$, but reinforcement of superconductivity was found around 15~T, where $M\simeq M_s$ [\onlinecite{aoki09}]. As in heavy fermions close to an antiferromagnetic-to-paramagnetic quantum instability \cite{aoki12b}, the moment polarization field in UCoGe and URhGe under $\mathbf{H}\parallel\mathbf{b}$ is connected to $T_\chi^{max}$ by a phase boundary (or crossover) in the $(H,T)$ plane and by a simple correspondence 1~K~$\leftrightarrow1$~T. For $\mathbf{H}\parallel\mathbf{b}$, contrary to URhGe where $T_C=T_{\chi}^{max}$, UCoGe is
\begin{table}[t]
\caption{Curie temperature $T_C$, spontaneous moment $M_s$, superconductive temperature $T_{sc}$ at $H=0$ and, for $\mathbf{H}\parallel\mathbf{b}$, temperature $T_{\chi}^{max}$ at the maximum of the magnetic susceptibility, moment polarization field $H_m$, and maximum $T_{sc}^{max}$ of the reentrant superconductive temperature when $H>H_{c,2}$, in URhGe and UCoGe.}
\begin{ruledtabular}
\begin{tabular}{lcc}
&URhGe&UCoGe\\
\hline
$T_C$ (K)&9.5&3\\
$M_s$ ($\mu_B$)&0.4&0.05\\
$T_{sc}$ ($H=0$) (K)&\;\,0.25&0.6\\
$T_{\chi}^{max}$ [$\mathbf{H}\parallel\mathbf{b}$] (K)&10&37.5\\
$\mu_0H_{m}$ [$\mathbf{H}\parallel\mathbf{b}$] (T)&12&50\\
$T_{sc}^{max}$ [$H>H_{c,2}$,$\mathbf{H}\parallel\mathbf{b}$] (K)&0.6&- $^\dag$\\
\hline
Refs.&[\onlinecite{aoki01,levy05}]&[\onlinecite{huy07}]\\
\end{tabular}
\end{ruledtabular}
 $^\dag$ In UCoGe, a $S$-shape of $H_{c,2}$ is observed at 10-15~T \cite{aoki09}.
\label{table}
\normalsize
\end{table}
characterized by a decoupling between the Curie temperature $T_C=3$~K and $T_{\chi}^{max}=37.5$~K. For UCoGe, the question is whether $T_\chi^{max}$ and $H_{m}$ are controlled by magnetic fluctuations or Fermi surface effects. The effect of a high magnetic field on the Fermi surface of UCoGe, which is a low-carrier system with large effective masses \cite{aoki11}, are expected to be important since the Zeeman energy easily affects the effective Fermi energy. Indeed, high-field Fermi surface reconstructions were reported in the low-carrier system URu$_2$Si$_2$ [\onlinecite{scheerer12,aoki12c}]. Theoretical calculations have shown that strong differences are expected between the Fermi surface of the paramagnetic and ferromagnetic states of UCoGe [\onlinecite{samsel10}]. The 30-K temperature scale in UCoGe is also responsible for a shoulder in the nuclear spin-lattice relaxation rate versus temperature (at zero-field \cite{ohta10} and in a field of 2~T applied along $\mathbf{a}$, $\mathbf{b}$, and $\mathbf{c}$ \onlinecite{ihara10}) and by an enhancement of the magnetic fluctuations at the reciprocal lattice zone-center \cite{stock11}. Further investigations of both the magnetic fluctuations and the Fermi surface of UCoGe in high field are needed to determine what drives the field-induced moment polarization at $\mu_0H_m\simeq50$~T for $\mathbf{H}\parallel\mathbf{b}$. The description of the kink at $\mu_0H^*\simeq25$~T for $\mathbf{H}\parallel\mathbf{c}$ would also merit consideration. Knowing that a field-induced moment at the Co sites was reported from polarized neutron scattering in $\mu_0\mathbf{H}\parallel\mathbf{c}=12$~T [\onlinecite{prokes10}], the magnetic properties of UCoGe are expected to be more complex than that of URhGe.

To conclude, we have reported new high-magnetic-field magnetization and magnetoresistivity measurements on the ferromagnetic superconductors UCoGe and URhGe. For $\mathbf{H}\parallel\mathbf{b}$, a field-induced moment polarization in UCoGe at 50~T vanishes at temperatures higher than 30-40~K, which corresponds to a maximum in the magnetic susceptibility. Strong similarities were emphasized with the moment reorientation occurring in URhGe, also for $\mathbf{H}\parallel\mathbf{b}$, at 12~T. In both systems, despite very different spontaneous moments similar values of the magnetization were found to be reached in a high magnetic field.

We acknowledge J. B\'{e}ard, L. Bendichou, P. Delescluse, T. Domps, J-M Lagarrigue, J.-P. Nicolin, and T. Schiavo for technical support. This work was supported by Euromagnet II via the EU under Contract No. RII3-CT-2004-506239, and by the ERC Starting Grant NewHeavyFermion.


\begin{thebibliography}{20}


\bibitem{aoki12b} D. Aoki, W. Knafo, and I. Sheikin, arXiv:1204.5128.

\bibitem{sato01} M. Sato, Y. Koike, S. Katano, N. Metoki, H. Kadowaki, and S. Kawarazaki, J. Phys. Soc. Jpn. Suppl. A {\bf70}, 118 (2001).

\bibitem{raymond99} S. Raymond, D. Raoelison, S. Kambe, L. Regnault, B. F{\aa}k, R. Calemczuk, J. Flouquet, P. Haen, and P. Lejay, Physica B {\bf259-261}, 48 (1999).

\bibitem{flouquet04} J. Flouquet, Y. Haga, P. Haen, D. Braithwaite, G. Knebel, S. Raymond, and S. Kambe, Physica B {\bf272-276}, 27 (2004).

\bibitem{knafo09} W. Knafo, S. Raymond, P. Lejay, and J. Flouquet, Nat. Phys. {\bf5}, 753 (2009).

\bibitem{aoki01} D. Aoki, A. Huxley, E. Ressouche, D. Braithwaite, J. Flouquet, J.-P. Brison, E. Lhotel, and C. Paulsen, Nature {\bf413}, 613 (2001).

\bibitem{levy05} F. L\'evy, I. Sheikin, B. Grenier, and A.D. Huxley, Science {\bf309}, 1343 (2005).

\bibitem{miyake09} A. Miyake, D. Aoki, and J. Flouquet, J. Phys. Soc. Jpn. {\bf78}, 063703 (2009).

\bibitem{huy07} N.T. Huy, A. Gasparini, D.E. de Nijs, Y. Huang, J.C.P. Klaasse, T. Gortenmulder, A. de Visser, A. Hamann, T. G\"{o}rlach, and H.v. L\"{o}hneysen, Phys. Rev. Lett. {\bf99}, 067006 (2007).

\bibitem{aoki09} D. Aoki, T.D. Matsuda, V. Taufour, E. Hassinger, G. Knebel, and J. Flouquet, J. Phys. Soc. Jpn. {\bf78}, 113709 (2009).

\bibitem{saxena00} S.S. Saxena, P. Agarwal, K. Ahilan, F. M. Grosche, R.K.W. Haselwimmer, M.J. Steiner, E. Pugh, I.R. Walker, S.R. Julian, P. Monthoux, G.G. Lonzarich, A. Huxley, I. Sheikin, D. Braithwaite, and J. Flouquet, Nature {\bf406}, 587 (2000).

\bibitem{sheikin01} I. Sheikin, A. Huxley, D. Braithwaite, J. P. Brison, S. Watanabe, K. Miyake, and J. Flouquet, Phys. Rev. B {\bf64}, 220503 (2001).

\bibitem{yelland11} E.A. Yelland, J.M. Barraclough, W. Wang, K.V. Kamenev, and A.D. Huxley, Nature Physics {\bf7}, 890 (2011).

\bibitem{malone12} L. Malone, L. Howald, A. Pourret, D. Aoki, V. Taufour, G. Knebel, and J. Flouquet, Phys. Rev. B {\bf85}, 024526 (2012).

\bibitem{huy08} N.T. Huy, D.E. de Nijs, Y.K. Huang, and A. de Visser, Phys. Rev. Lett. {\bf100}, 077002 (2008).

\bibitem{noteURhGe} Due to a misalignment of the URhGe crystal measured by resistivity, the field at the moment reorientation is a bit higher than that reported usually \cite{levy05}.

\bibitem{BealMonod1982} M. B\'{e}al-Monod, Physica B+C {\bf109-110}, 1837 (1982).

\bibitem{aoki11} D. Aoki, I. Sheikin, T.D. Tatsuma, V. Taufour, G. Knebel, and J. Flouquet, J. Phys. Soc. Jpn. {\bf80}, 013705 (2011).

\bibitem{samsel10} M. Samsel-Czeka{\l}a, S. Elgazzar, P.M. Oppeneer, E. Talik, W. Walerczyk, and R. Tro\'{c}, J. Phys.: Condens. Matter {\bf22}, 015503 (2010).

\bibitem{scheerer12} G.W. Scheerer, W. Knafo, D. Aoki, G. Ballon, A. Mari, D. Vignolles, and J. Flouquet, Phys. Rev. B {\bf85}, 094402 (2012).

\bibitem{aoki12c} D. Aoki, G. Knebel, I. Sheikin, E. Hassinger, L. Malone, T.D. Tatsuma, and J. Flouquet, J. Phys. Soc. Jpn. {\bf81}, 074715 (2012).

\bibitem{ohta10} T. Ohta, T. Hattori, K. Ishida, Y. Nakai, E. Osaki, K. Deguchi, N.K. Sato, I. Satoh, J. Phys. Soc. Jpn. {\bf79}, 023707 (2010).

\bibitem{ihara10} Y. Ihara, T. Hattori, K. Ishida, Y. Nakai, E. Osaki, K. Deguchi, N.K. Sato, I. Satoh, Phys. Rev. Lett. {\bf105}, 206403 (2010).

\bibitem{stock11} C. Stock, D.A. Sokolov, P. Bourges, P.H. Tobash, K. Gofryk, F. Ronning, E.D. Bauer, K.C Rule, and A.D. Huxley, Phys. Rev. Lett. {\bf107}, 187202 (2011).

\bibitem{prokes10} K. Prokes, A. de Visser, Y.K. Huang, B. F{\aa}k, and E. Ressouche, Phys. Rev. B {\bf81} 180407(R), (2010).





































\end{thebibliography}
\end{document}